\documentstyle[12pt,epsf]{article}   
\newcommand{\frn}[2]{\frac{\displaystyle#1}{\displaystyle#2}}
\font\am      =msam10 scaled 1440
\newcommand{\ale}{\mathbin{\mbox{\am\symbol{'056}}}}
\newcommand{\age}{\mathbin{\mbox{\am\symbol{'046}}}}

\renewcommand{\ge}{\;\mbox{\am\symbol{'076}}\;} 
\newcommand{\eps}{\varepsilon}
\newcommand{\mn}{\mu \nu}
\newcommand{\ds}{\displaystyle}

\newcommand{\bel}[1]{\begin{equation}\label{#1}}
\newcommand{\ee}{\end{equation}}

\begin{document}

\begin{titlepage}   
   
\begin{flushright}   
Preprint MSU Faculty of Physics 7/98
\end{flushright}   
\bigskip   
   
\begin{center}   
\large\bf   
Production of Higgs with Z--boson by an electron
in external fields
\end{center}   
\vspace{0.5cm}   
\begin{center}   
P. A. Eminov\\
{\sl 
Department of Physics\\
Moscow State Institute of Electronics and Mathematics\\
(Technical University)\\
109028, Moscow, Russia}\\[15pt]
K. V. Zhukovskii, K. G. Levtchenko \\
{\sl 
Faculty of Physics,\\
Department of Theoretical Physics, Moscow State University,\\ 
119899, Moscow, Russia}   
\end{center}   
 
\begin{abstract}   
The rate of associative production of Higgs and $Z$--bosons by charged
leptons in the field of a plane electromagnetic wave of arbitrary intensity 
and in the constant crossed field is obtained.
The cross section is examined as a function of particle energies and external
field intensities for various values of the Higgs boson mass.
It is shown, that the photoproduction cross section increases logarithmically
at super high energies up to the values, that essentially exceed the cross
section of the reaction $e^+e^-\to ZH,$ which, at present, is considered as the
most probable channel of Higgs boson production.
\end{abstract}   
   
\vspace{0.5cm}   
  
\vfill   
\end{titlepage}   

\noindent 
\section{ Introduction} 
 
The Higgs mechanism of spontaneous symmetry breaking is 
one of the key elements in the electroweak sector of the Standard 
Model, along with the principle of gauge invariance. It is due to this principle that 
fundamental particles, quarks and weak gauge bosons, acquire 
masses 
through their interaction with a scalar Higgs field. 
 
At present, the fundamental massive Higgs boson is the only 
particle of the Standard Model which has not been observed so 
far. Experimental discovery of the scalar Higgs bosons could provide an 
important test of the Standard Model and even of the Higgs 
mechanism of spontaneous symmetry breaking in particle 
physics itself.  
 
According to the Weinberg--Salam--Glashow (WSG) theory the 
masses of 
the $W^\pm$-- and $Z$--bosons as well as the vacuum expectation value 
$v$ of 
the Higgs field can be written 
in terms of the Fermi constant $G_F$, the fine structure constant $\alpha,$  and  the Weinberg 
angle $\theta_W$ [1, 2]: 
$$ 
M_W=\left(\frn{\pi\alpha}{\sqrt{2}G_F}\right)^{1/2}\frn{1}{\sin\theta_W} 
=80,37 \pm 0,19\,Gev, 
$$ 
$$ 
M_Z=\frn{M_W}{\cos\theta_W}=91186,3 \pm 1,9\,Mev,\;
v=(\sqrt{2 }G_F)^{-1/2}\simeq246\,Gev.
$$ 
 
The vacuum expectation value $v$ and the weak mixing angle $\theta$ are 
experimentally determined, so that the only unknown parameter is 
the mass of the Higgs particle: 
$$ 
M_H=\lambda  v, 
$$ 
which depends on the as yet unknown Higgs boson self--coupling constant $\lambda.$ 

Since in the WSG theory couplings of Higgs bosons to other particles 
grow with the masses of these particles, interactions of 
Higgs bosons with gauge bosons and heavy quarks are much stronger than those
with electrons and other light particles. Therefore processes 
of associative production of Higgs particles with gauge $W^\pm$-- and $Z$--bosons are 
considered to be the most 
effective ones in the entire range of possible Higgs production 
mechanisms and are of substantial theoretical and practical interest. 
 
The main Higgs particle production mechanisms in $e^+e^-$--collisions 
are the Higgs strahlung off the $Z$--boson line $e^++e^-\to Z\to Z+H$ 
and the fusion processes $e^++e^-\to W^+W^-\nu_e\tilde\nu_e\to
H+\nu_e+\tilde\nu_e,$ $e^++e^-\to e^+e^-ZZ\to e^++e^-+H$ [3--6], 
among which the $e^++e^-\to Z+H$ process dominates at    
$\sqrt{s}\ale 500\,$GeV [7],  where
$\sqrt{s}$  is the c.m. energy of colliding particles. The direct search 
for the process $e^++e^-\to Z+H$ at LEP2 sets a lower
bound  for the Higgs mass of $M_H >75$\,GeV [1]. In this case the cross  
section does not exceed $0.3\,$pb for the
Higgs boson mass $M_H \in (50,350)$\,GeV, and it decreases at higher values of $M_H$.

Taking into account  
the direct measurements of $t$--quark and $W$--boson masses at Tevatron,
the mass of the Higgs particle is considered to be $\ds 127^{+127}_{-72}$\,GeV,
and at a confidence level of 95\,\%  it must be less than $465$\,GeV [1].  
 
As for the upper limit for the $H$--boson mass in the Standard Model, it can
be derived from the estimation of 
the energy range within which the model is
assumed to be valid, i.e., before the particles interaction becomes strong. 
Taking
into account that the strength of the Higgs self--interaction as well as  of 
the
interaction of $W$-- and $Z$--bosons
with Higgs particle is determined by the
Higgs mass and the constant $\lambda$, we obtain that at 
$M_H \gg M_Z,$ $M_H \gg M_W$ interaction between particles appears to be strong. 
The detailed analysis
leads to an estimate of about 700 GeV for the upper limit of $M_H$ [7--10].

Electron--photon collisions are among other possible channels of Higgs boson production.
For example, the dependence  of the cross section of the process
$e+\gamma \to W+H+\nu_e$ upon the mass of the Higgs boson for the  
energy range $\sqrt{s}=200\div 2000$\,GeV is studied in [11, 12] and as 
it was shown in [12],  production of Higgs bosons with mass values greater
than 140 GeV in the reaction $e+\gamma \to e\gamma \gamma \to e+H,$   is quite feasable
for $\sqrt{s}>500$\,GeV.
To produce hard photons in this case the inverse Compton effect can be used
with a nearly monochromatic spectrum of scattered photons for
$\chi=\frn{2\omega E}{m^2} \gg1$, having a sharp maximum at
$\omega'\approx E$. Here $\omega,$ $\omega',$ $E$ are the energies of  
the incident and scattered photons, and of the relativistic electron respectively.  
 
	The Higgs bosons production and decay processes in the 
presence of external electromagnetic fields are actually of great interest.
The importance of these studies are  determined by the fact that
in the field of an external electromagnetic wave the rates of many processes, that are 
usually forbidden by the 4-momentum conservation law, increase substantially, 
reaching measur\-able values, and, on the other hand,
many reactions become more informative in the  
presence of external electromagnetic fields [13--15]. 

	 In the present paper the associative production of Higgs 
and $Z$--bosons by  
a charged lepton in external electromagnetic fields of various 
con\-fi\-gu\-ra\-tions is examined. The rates are calculated with the use of the relativistic wave 
functions of 
the particles in the external field which enables interactions of charged particles 
with external electromagnetic field to be considered exactly [13--16].

	 In Section 2 the rate of the process $e\to e+Z+H$ in the 
presence of the electromagnetic  wave field of arbitrary intensity and
configuration is calculated. 

	In Section 3 the crossed field case is examined. Under 
the condition, that the  
electron is extremely relativistic and the intensity of the field is relatively 
small ($E,\,H \ll H_0=\frn{m^2}{e}
\simeq 4,41\cdot 10^{13}$ Gs), the results obtained  
can be applied to the case of
an arbitrary constant field. 

	 In Section 4 the asymptotics for the Higgs boson production rate in the cases of the 
crossed field and an electromagnetic wave are obtained. 
It is shown that the Higgs  production cross section in the reaction $e+\gamma \to e+Z+H$ with 
photons absorbed from the wave and with high energy electrons can exceed  
the rate of the reaction $e^-+e^+\to Z+H$, which is considered to be one of 
the most probable channel of Higgs production [1, 3, 4].

\noindent
\section{Process $e\to e+Z+H$ in the presence of an external 
elec\-tro\-mag\-netic wave.}
  
	In the  Standard Model framework the
matrix  element of the process examined can be 
written as follows [17]  
$$
\langle f|S^{(2)}|i\rangle=\frn{ig^2M_Z}{\cos^2\Theta_W\sqrt{4k_0k'_0}}J^{\mu}
\left[g_{\mn}-\frn{p_{\mu}p_{\nu}}{M^2_Z}\right]
\frn{e^{(\lambda)*}_{\nu}(k')}{p^2-M^2_Z +i\Gamma_Z M_Z},
$$
where $\Gamma_Z\simeq 2494,7 \pm 2,6$ MeV is the $Z$--boson 
decay width , $k=(k_0,{\bf k}),$ $k'=(k'_0, {\bf k'}),$ $p$ are the 
momenta of the Higgs boson, outgoing $Z$--boson,
and virtual $Z$--boson respectively, and
$$
J^{\mu}=\int d^4x\bar\psi_{q'}(x)\gamma^{\mu}(g_V+g_A \gamma^5)
\psi_q(x)\exp(ipx)
$$
is the electroweak current. 
Here $g_A=-1/4,\; g_V=-1/4+\sin^2\Theta_W$ and $\psi_q(x)$ is the exact
solution of the  Dirac equation for an electron in the given external field.

The wave function of an electron in the field of an external plane electromagnetic wave  
described by the  4--potential $A^{\mu}=A^{\mu}(\varphi)$
(the phase $\varphi=nx$, where $n$ is a wave 
vector, $n^2=0$), is given by  the following 
expression [14,18]:
\bel{3}
\psi_q(x)=(2q_0V)^{-1/2} \left[1+\frn{e}{2(nq)}(\gamma n)(\gamma A) \right]
u(q)\exp(iS_q(x)).
\ee
Here $V$ is the normalization volume, and $u(q)$ is  the bispinor amplitude of
the free particle  solution of the Dirac equation
$$
(\gamma q- m ) u(q)=0,\;\; q^2=m^2,
$$
and $S_q(x)$ coincides with the classical action 
for the particle, moving in the  
external plane wave field: 
\bel{5}
S_q(x)=-qx-\int\limits^{\varphi}_0d\varphi
\left[\frn{e}{(nq)}(qA)-\frn{e^2A^2}{2(nq)}\right].
\ee

In the case of a circularly polarized plane wave, determined by the potential
$$
A^{\mu}(x)=a^{\mu}_1 \cos\varphi+ a^{\mu}_2 \sin\varphi,
$$
$$
a^2_1=a^2_2=a^2,\; a_1 a_2=0,\; a_1n=a_2n=0,\; \varphi=nx,
$$
the following expression
for the wave function is obtained from (\ref{3}) and (\ref{5}): 
$$
\psi_q(x)=\left[1+\frn{e}{2(nq)}(\hat n\hat a_1 \cos\varphi+
\hat n\hat a_2 \sin\varphi)\right]\times
$$
$$
\times\frn{u(q)}{\sqrt{2Q_0}}
\exp\left\{-ie\frn{a_1q}{n q}\sin\varphi+
ie\frn{a_2q}{n q}\cos\varphi-iQx\right\}.
$$
  
Here the electron momentum  in the presence of an external plane wave field 
was introduced
$$
Q^{\mu}=q^{\mu}-e^2\frn{a^2}{2(n q)}n^{\mu},
$$
whose square is equal to the electron effective mass in 
the presence of an external field: 
$$
Q^2=m^2_*=m^2(1+\xi^2).
$$
In this expression $\xi=\sqrt{-\frn{e^2a^2}{m^2}}$ is the
well known parameter of the wave intensity
determined by the ratio of the work produced by the field
in its wavelength to the electron energy at rest.
  
	The  averaging of the squared
matrix element with  
respect to the spin states of initial electron and summing over  
polarizations of the outgoing  
electron is carried out according to the common procedure, while the sum over 
polarizations of the $Z$--boson is obtained with the use of the following 
formula: 
$$
\sum_{\lambda=1,2,3}e^{(\lambda)}_{\nu}(k')e^{(\lambda)}_{\mu}(k')=
-\left(g_{\mn}-\frn{k'_{\mu}k'_{\nu}}{M^2_Z}\right),
$$
where $e^{(\lambda)}_{\nu}(k')$  is the $Z$--boson polarization  4--vector. 

	Upon integrating, made in the tensor form, with respect to the Higgs  
and the outgoing 
$Z$--boson phase volumes the rate of the  process 
in the unit space 
volume is obtained
$$
W=\frn{G^2_F}{(2\pi)^3}M^6_Z
\frn{m^2}{Q_0} \sum_{s>s_0}\int\limits^{u_2}_{u_1}\frn{du}{(1+u)^2}
\int\limits^{\tau(u)}_{M^2}
\frn{d\tau\sqrt{(\tau-M^2_Z-M^2_H)^2-4M^2_Z M^2_H}}{\tau((\tau-M^2_Z)^2+
(\Gamma_Z M_Z)^2)}\times
$$
\bel{11}
\times\left\{AE-4g^2_A \frn{m^2}{M^2_Z} F\left(B\frn{(\tau-M^2_Z)^2}{M^4_Z}+
A\left(2-\frn{\tau}{M^2_Z}\right)\right)\right\},
\ee
$$
E=(g^2_A +g^2_V)\left[-2\xi^2\frn{u^2+2u+2}{u+1}(J^2_{s+1}+J^2_{s-1}-2J^2_s)-
\right.
$$
\bel{12}
\left.-8J^2_s\left(1-\frn{\tau}{2m^2}\right)\right]+16(g^2_V-g^2_A)J^2_s+
\ee
$$
+2g_A g_V\frn{u+2}{u+1}J_s(J_{s-1}-J_{s+1})\frn{np}{m^2}
4z\left(1-\frn{2\xi^2}{1+\xi^2}\frn{u}{u_sz^2}\right),
u_s=\frn{2s(np)}{m^2_*},
$$
$$
F=-2\frn{\tau}{m^2}J^2_s+\frn{u^2}{u+1}\xi^2(J^2_{s+1}+J^2_{s-1}-2J^2_s),\;\;
M=M_Z + M_H,
$$
$$
A=\frn{8\tau M^2_Z+(\tau+M^2_Z - M^2_H)^2}{12\tau M^2_Z},\;\;
B=\frn{(\tau+M^2_Z - M^2_H)^2-\tau M^2_Z}{3\tau^2}.
$$
 
Each term in the series (\ref{11}) corresponds to the Higgs and 
$Z$--bosons production accompanied by absorption of $s$ photons from the wave,
with their minimal number being equal to  
$$
s_0=\frn{(M + m_*)^2-m^2_*}{2(nq)}.
$$
 
In the expressions (\ref{11}), (\ref{12}) for the rate of the process
the new variables
$u=-1+\frn{(nq)}{(nq')},$ $\tau=(sn+q-q')^2,$ and the boundaries 
$$
u_{1,2}=\frn{(sn+q)^2- M^2 - m^2_*
\pm \sqrt{((sn+q)^2- M^2 - m^2_*)^2- 4m^2_*M^2 }}{m^2_*},
$$
$$
\tau(u)=\frn{(sn+q)^2u}{1+u}- m^2_* u.
$$
were introduced.

The argument of the Bessel functions in the expression (\ref{12}) for the rate is equal to 
$$
z=2s\frn{\xi}{\sqrt{1+\xi^2}}
\sqrt{\frn{u}{u_s}\left(1-\frn{u}{u_s}- \frn{\tau(1+u)}{uu_s m^2_* }\right)}.
$$

	The result obtained (\ref{11}), (\ref{12}) is exact, i.e., it is valid for any value 
of the classical parameter of the wave intensity, the  nonlinearity domain, $\xi^2\age 1,$ 
including. In this region interaction of an 
electron with the
intensive electromagnetic wave field leads to new effects,
nonlinearly depending on the energy density of the wave.   

	Below a particular case, $\xi\ll 1,$ will be considered, when the 
perturbation theory can be
applied and the processes with minimal number of absorbed photons 
are most probable.  
Upon representing (\ref{11}), (\ref{12}) as a power series
in  $\xi^2,$ the following condition  
\bel{nq}
2(nq)>(M + m_*)^2-m^2_*,
\ee
is to be satisfied, which means that the process with a single photon 
absorbed from the wave becomes possible.

 Thus, dividing the rate (\ref{11}) by the incident current density
$j=\frn{m^2 \ae}{2\omega EV}$ ($\omega$ is the photon energy, $E$ is the electron energy, 
$\ae=\frn{2(nq)}{m^2}$) and putting $\xi^2=\frn{4\pi\alpha}{m^2 \omega V}$ ($\alpha$ is the 
fine structure constant), the
cross section of the process $e+\gamma\to e+Z+H$ is  finally obtained    
$$
\sigma=\left(\frn{eG_F m}{\pi}\right)^2\left(\frn{M^2_Z}{\ae m^2}\right)^2
\int\limits^{1-a}_b\frn{d \lambda (1- M^2 /(\ae\lambda m^2))^{1/2}
(1-M^2_1 / (\ae\lambda m^2))^{1/2}}{(\lambda - M^2_Z / (\ae M^2))^2}\times
$$
\bel{sig}
\times\left\{2A C -4g^2_A \frn{m^2}{M^2_Z}D\left[B\frn{m^4 \ae^2}{M^4_Z}
\left(\lambda - \frn{M^2_Z}{\ae m^2}\right)^2+ A \left(2-
\frn{\ae\lambda m^2}{M^2_Z}\right)\right]\right\},
\ee
where $\lambda=\frn{\tau}{m^2 \ae},$
$$
C=(g^2_V+g^2_A)[2\lambda(1-\lambda)-1]\ln\frn{1-\lambda}{a}-
2\lambda(g^2_V+g^2_A)(1-\lambda- a) +
$$
$$
+4 g_Vg_A \left(\frn{1}{2}-\lambda \right)\ln\frn{1-\lambda}{a}-
4g_Vg_A (1-\lambda -a ),\;\;
M_1=M_Z-M_H,
$$
$$
D=[1-2\lambda(1-\lambda)]\ln\frn{1-\lambda}{a}+2\lambda(1-\lambda- a ),\;\;
b=\ae a=\frn{1}{\ae}\frn{M^2}{m^2},
$$
and $A, B$ are determined by (\ref{12}).

\noindent
\section{$e\to e+Z+H$ process in the presence of a constant crossed field.} 
 
	In the present section production of  
Higgs bosons in constant  
crossed fields with intensity vectors ${\bf E}$ and ${\bf H}$ normal 
and with their magnitudes equal $|{\bf E}|=|{\bf H}|$ to each other (this  
means that both of the field invariants are equal to zero) is considered.

This crossed field can be regarded as an exceptional limiting case of a plane wave 
electromagnetic field, determined by the vector potential
\bel{pot}
A^{\mu}=a^{\mu}\varphi,\;\; an=0.
\ee
Thus, with regard to (\ref{3}) there follows an expression for the electron 
wave function in the crossed field 
    
$$
\psi_q(x)=\left[1+\frn{e(\gamma n)(\gamma a)}{2(nq)}\varphi\right]
\frn{u(q)}{\sqrt{2q_0V}}\times
$$
$$
\times\exp\left[-ie\frn{aq}{2(nq)}\varphi^2+
ie^2a^2\frn{\varphi^3}{6(nq)}-iqx\right].
$$
 
	The rate of the process $e\to e+Z+H$ can be derived from (\ref{3}) 
in a standard way after 
some computations
with the use of the  electron wave function in a crossed field, but here an 
alternative  method for 
calculating the rate, based on the  exact result (\ref{11}) obtained
for the case of a circulary polarized wave will be employed. Indeed, the total rate of the process 
$e\to e+Z+H$ in the presence of such a wave depends on the two  
invariant parameters 
$$
\xi=\sqrt{-\frn{e^2a^2}{m^2}}=\frn{eF}{m \omega},\;\;
\chi=\frn{e}{m^3}[-(F^{\alpha \beta}q_{\beta})]^{1/2}=\xi\frn{(nq)}{m^2}.
$$

Herewith the electric and magnetic field intensity vectors
rotate in the plane normal to the direction of the wave propagation
at the frequency equal to the frequency of the wave.
 
	Thus, the total rate of the process, 
calculated
in the presence of a plane wave with circle polarization for 
$\omega \to0\; (\xi\to\infty)$ 
and that in a constant crossed field must be exactly equal [14, 15, 18]: 
\bel{t2}
\lim_{\xi\to\infty}W(\xi,\chi)\equiv W(\infty,\chi)\equiv W(\chi).
\ee
 
It should be noted, that the result 
obtained in the limit (\ref{t2}) is valid for the crossed electromagnetic field at arbitrary value
of the electron energy, and moreover, if $\eps\gg m,$ (i.e. extremely relativistic
case) it describes the rate
of the process under consideration for an arbitrary configuration of a constant relatively weak 
electromagnetic field $F\ll H_0$ [14].

Next, the order of summing in
the series and integrating in (\ref{11}) are interchanged yielding the expression  
$$
W=\sum_{s>s_0}\int\limits^{2\pi}_0d\varphi\int\limits^{\infty}_0du
\int\limits^{\infty}_{M^2}d\tau W(u,\tau,s,\varphi)=
$$
$$ 
=\int\limits^{2\pi}_0d\varphi\int\limits^{\infty}_0du
\int\limits^{\infty}_{\left(\frac Mm \right)^2} d\alpha
\sum_{s>s_{min}}m^2W(u,\alpha,s,\varphi),
$$
where 
$$
\alpha=\frn{\tau}{m^2},\; s_{min}=\frn{\xi^3u}{2\chi}\left[1+\frn{1}{\xi^2}\left(1+\alpha
\frn{u+1}{u^2}\right)\right],\;\;
\chi=\frn{nq}{m^2}\xi.
$$
 
If $\xi^2\gg1,$ the value of the rate of the process is determined by 
the region   $z\sim s\sim \xi^3 \gg1,$  since it gives the dominant
contribution to the result. Therefore, we can substitute summation
over $s$ by integration with respect to a new variable $\tau$ 
using the following relations between s and $\tau$: 
$$
s=\frn{\xi^3 u}{2\chi}\left(1+\frn{2\tau}{\xi}\right)+s_{min}.
$$
 
The final result is 
\bel{t4}
W=\int\limits^{2\pi}_0d\varphi\int\limits^{\infty}_0du
\int\limits^{\infty}_{\left(\frac Mm \right)^2} d \alpha
\int\limits^{\infty}_{-\frac \xi2}d\tau\frn{\xi^2u}{\chi} m^2
W(\varphi,u,\tau,\alpha).
\ee

	For  high values $\xi^2\gg1$  the rate 
$W(\varphi,u,\tau,\alpha)$ can be representd  using the following asymptotics 
[19] for the Bessel
functions, which is valid when their argument and index tend to infinity, while their
ratio tends to unity, : 
$$
J_s(z)\simeq\frn{1}{\pi}\left(\frn{2}{s}\right)^{1/3}\Phi(y),
$$
where $\Phi(y)$ is the Airy function of  the argument  
$$
y=\left(\frn{s}{2}\right)^{2/3}\left(1-\frn{z^2}{s^2}\right)=
\left(\frn{u}{2\chi}\right)^{2/3}\left[1+\alpha \frn{u+1}{u^2}+\tau^2\right].
$$

After the limiting procedure has been performed, integration in (\ref{t4}) with respect to the 
angular variable $\tau$ be can carried out 
taking into account the well known relations for the Airy functions, cited in  [14].
Thus,  the following representation for the total rate of the process 
$e\to e+Z+H$ in the presence
of  a crossed field  is finally obtained
\bel{t5}
W=-\frn{1}{\sqrt{\pi}}\frn{G^2_F M^6_Z}{(2\pi)^3Q_0}
\left(\frn{m}{M}\right)^2\int\limits^1_0
\frn{dx}{\left(1-x\frn{M^2_Z}{M^2}\right)^2}\times
\ee
$$
\times(1-x)^{1/2}\left(1-x\frn{M^2_1}{M^2}\right)^{1/2}
\int\limits^{\infty}_0\frn{du}{(u+1)^2}G(u,x),
$$
where 
$$
G(u,x)=4F_1G_1-8g^2_A \frn{m^2}{M^2_Z}G_2\left[\frn{F_2}{x^2}
\left(1-x\frn{M^2_Z}{M^2}\right)^2\left(\frn{M}{M_Z}\right)^4+\right.
$$
$$
\left.+\frn{F_1}{x}\left(\frn{M}{M_Z}\right)^2
\left(2x\frn{M^2_Z}{M^2}-1\right)\right],
$$
$$
G_1=4(g^2_A -g^2_V)\Phi_1+(g^2_A +g^2_V)
\left[(2-\left(\frn{M}{m}\right)^2x)\Phi_1-2\frn{u^2+2u+2}{u+1}\Phi'
\left(\frn{\chi}{u}\right)^{2/3}\right],
$$
$$
G_2=\left(\frn{M}{m}\right)^2x \Phi_1+2\frn{u^2}{u+1}\Phi'\left(\frn{\chi}{u}\right)^{2/3},
$$
\bel{t6}
F_1=\frn{2}{3}+\frn{1}{6}\left(1-\frn{M^2_H}{M^2_Z}\right)+
\frn{1}{12}\left(\frn{M}{M_Z}\right)^2\frn{1}{x}+
\frn{1}{12}\left(1-\frn{M^2_H}{M^2_Z}\right)^2\left(\frn{M_Z}{M}\right)^2x,
\ee
$$
F_2=\frn{1}{3}+\frn{2}{3}\frn{M^2_Z - M^2_H}{M^2}x+
\frn{(M^2_Z - M^2_H)^2}{3M^4}x^2 - \frn{1}{3}\left(\frn{M_Z}{M}\right)^2x,
$$
In Eq. (\ref{t6}) the function $\Phi_1(z)$ is defined as   
$$
\Phi_1(z) =\int\limits^{\infty}_z\Phi(t)dt,
$$
where the argument $z$ is equal to  
\bel{t7}
z=\left(\frn{u}{\chi}\right)^{2/3}\left[1+\left(\frn{M}{m}\right)^2\frn{1}{x}
\frn{u+1}{u^2}\right].
\ee

In the case of an extremely relativistic
electron with the energy $E\gg m,$ and momentum component $p_Z=0,$ in 
a constant magnetic field  
${\bf H}\,\uparrow\uparrow\,Oz$ of relatively low intensity  
$H\ll H_0=\frac{m^2}{e}=4,41\cdot10^{13}$ Gs
the spectral variable $u$ and the dynamic parameter $\chi$ in
(\ref{t5})--(\ref{t7}) are defined as 
$$
u=\frac{p_{\perp}}{p'_{\perp}}-1=\sqrt{\frac{n}{n'}}-1,\; \chi=\frac{H}{H_0}
\frac{p_{\perp}}{m},
$$
where $p_{\perp}=\sqrt{2eHn}$ is the value of the transversal component of 
the electron momentum in the 
magnetic 
field, $n$ is the principal quantum number, and the
electron energy spectrum is given by the 
following expression  [16]:
$$
E=\sqrt{2eHn+ m^2 +p^2_Z}.
$$

\noindent
{\section{The limiting cases and the discussion of the results obtained.}

	Some of the interesting features of the results 
obtained concerning the Higgs boson production in the presence 
of a constant crossed field are to be discussed in more detail:
 
	a)  When the dynamic parameter is relatively small, 
$\chi\ll \left(\frac{M}{m}\right)^2,$ the rate (\ref{t5}) 
is mainly formed  in the region $z\gg1$, where the following asymptotics  
for the Airy function
is valid: 
\bel{th24}
\Phi(z)\simeq\frac12z^{-1/4}\exp\left(-\frac23z^{3/2}\right).
\ee

With regard to (\ref{th24}) integration of (\ref{t5}) with respect to the
spectral variable can be carried out by means of the saddle-point method, where the saddle-point 
$u_0$  is derived as a solution of the equation  
$$
2-\frac{\lambda}{u_0}-4\frac{\lambda}{u^2_0}=0,\;\; 
\lambda=\left(\frn{M}{m}\right)^2\frn{1}{x}, 
$$
which leads to $u_0\simeq\frac{\lambda}2\gg1.$ 

As a result, we represent (\ref{t5}) as a simple integral with respect to 
$\lambda$:
\bel{th25}
W=\frac{G^2_F M^6_Z}{(2\pi)^3Q_0}\frac{16}{\sqrt{3}}\,\chi
\int\limits^{\infty}_{\left(\frac Mm \right)^2}
\frn{d\lambda}{\lambda}G(\lambda)
\left[1-\frac{M^2}{m^2 \lambda}\right]^{1/2}
\exp\left[-\sqrt{3}\frac{\lambda}{\chi}\right],
\ee
where 
$$
G(\lambda)=\left[1-\frac{M^2_1}{m^2 \lambda}\right]^{1/2}
\frn{1}{\left(\lambda -\frn{M^2_Z}{m^2}\right)^2}\times
$$
$$
\times\left\{(g^2_V+g^2_A)F_1+2g^2_A \left(\frac{m}{M_Z}\right)^2
\left[F_2\frac{m^4}{M^4_Z}\left(\lambda -\frac{M^2_Z}{m^2}\right)^2+
F_1\left(2-\lambda \frac{m^2}{M^2_Z}\right)\right]\right\}.
$$

Applying the saddle-point method to 
(\ref{th25}) again,  the Higgs production rate can  finally be obtained. 
For relatively small values of 
the dynamic parameter 
$\chi\ll \left(\frac{M}{m}\right)^2$ it has the form: 
$$
W\simeq\frac{8G^2_F M^6_Z}{(2\pi)^3Q_0}
\frn{\sqrt{2\pi}}{\psi^{5/2}}\left(\frn{M}{m} \right)^2
G\left(\frac{M^2}{m^2}\right)\exp(-\psi),
$$
$$
\psi=\sqrt{3}\left(\frac{M}{m}\right)^2\frac1{\chi}.
$$
In this case the value of the function $G(\lambda)$ is  taken at the saddle-point
$\lambda=\left(\frn{M}{m}\right)^2$.
 
	It should be noted that the exponential behavior of 
the rate in the case of  relatively small values of the parameter $\chi$ is 
characteristic for the 
processes forbidden in the absence of an external field. 

    b)  Below we shall calculate the Higgs
production rate in the most important case of relatively high values of the dynamic 
parameter  $\chi\gg\left(\frac{M}{m}\right)^2.$ Then the argument
of Airy function in the region that provides the largest contribution to rate 
(\ref{t7}), becomes: 
\bel{th28}
z\simeq\left(\frac{M}{m}\right)^2\frac1{\chi^{2/3}u^{1/3}x}.
\ee
 
In contrast to the case $\chi\ll \left(\frac{M}{m}\right)^2,$ when the value
of the integral over
the spectral variable is determined mainly by the saddle-point neighborhood
$u_0\simeq\frac{\lambda}2\ge\frac12\left(\frac{M}{m}\right)^2\gg1,$ 
in the case under consideration $\chi\gg\left(\frac{M}{m}\right)^2$ the 
main  contribution to the integral is provided by 
the relatively
wide domain $1 \ll u \ll \left(\frac{M}{m}\right)^2$.

Integration with respect to variable  $u$ with regard to (\ref{th28}) is carried out
by means of the integrals: 
$$
\int\limits^{\infty}_0t\Phi'(t)dt=-\Phi_1(0)=-\frn{\sqrt{\pi}}{3},
$$
$$
\int\limits^{\infty}_0t^2\Phi_1(t)dt=\frac23\Phi_1(0).
$$
  
Thus, for the rate of  the reaction $e\to e+Z+H$ we finally obtain: 
\bel{th29}
W=-\frac{16G^2_F M^6_Z}{3(2\pi)^3Q_0}\left(\frac{m}{M}\right)^6\chi^2\times
\ee
$$
\times\int\limits^1_0\frac{x^2dx}{\left(1-x\frac{M^2_Z}{M^2}\right)^2}
(1-x)^{1/2}\left(1-x\frac{M^2_1}{M^2}\right)^{1/2}\times
$$
$$
\times\left\{(g^2_V+g^2_A)F_1+2g^2_A \left(\frac{m}{M_Z}\right)^2
\left[\frac{F_2}{x^2}\left(1-x\frac{M^2_Z}{M^2}\right)^2
\left(\frac{M}{M_Z}\right)^4+\right.\right.
$$
$$
\left.\left.+\frac{F_1}{x}\left(\frac{M}{M_Z}\right)^2
\left(2x\frac{M^2_Z}{M^2}-1\right)\right]\right\}.
$$
 
	The integral with respect to $x$  in (\ref{th29}) can be easily
calculated, but the complete expression obtained is too complicated to be presented
here. The asymptotics of the rate (\ref{th29}) in the limiting cases, $M_H \gg M_Z$  and
$M_H=M_Z,$ have the following form: 
\bel{th30}
\hspace*{3em}W=C(g^2_V+g^2_A)\left\{\ds \frac1{240}
\left(\frac{M_H}{M_Z}\right)^2, \;\;\; M_H \gg M_Z, \atop
\ds\frac{16(23\sqrt{3}\pi-125)}9, \;\;\; M_H=M_Z, \right.
\ee
where  $\ds C=\frac{16}{3}\frac{G^2_F M^6_Z}{(2\pi)^3Q_0}
\left(\frac{m}{M}\right)^6\chi^2.$

If the mass of the Higgs particle is large enough $M_H \gg M_z,$ the result 
(\ref{th30}) up to a numerical factor coincides with
that, obtained in [20], where the rate of the process $e\to e+Z+H$ in the
superstrong magnetic field was calculated. It was shown there, that the
associative  Higgs and $Z$ -- boson production is a fairly probable
process, at least in the presence of a superstrong magnetic field. 

 	Finally some interesting cases of the 
photoproduction process $e+\gamma \to e+Z+H$ are to be studied.

In the Fig. 1 the rate of the process $e+\gamma \to e+Z+H$ is depicted as a function of the 
parameter $\ae,$ calculated by means of the formulas (\ref{sig})
 in the intermediate range of the Higgs boson masses:
$M_H=100\,(1),$ $200\,(2)$ $300\,(3)$ $400\,Gev\,(4).$ Taking (\ref{nq}) into
account, we conclude that near the threshold, when $\ae\ge M^2 / m^2 \approx 
10^{11}\div 10^{12}\,$Gev, the cross section of the reaction
$e+\gamma \to e+Z+H$ is small
as compared with that of $e^-+e^+\to Z+H$.  
However, if $\sqrt{s}\gg M_H$ (where $\sqrt{s}$ is the c.m. energy of the
particles) , the cross section of the process $e^-+e^+\to Z+H$
decreases as $s^{-1},$ according to [17] 
\begin{figure}
\centerline{\epsfbox{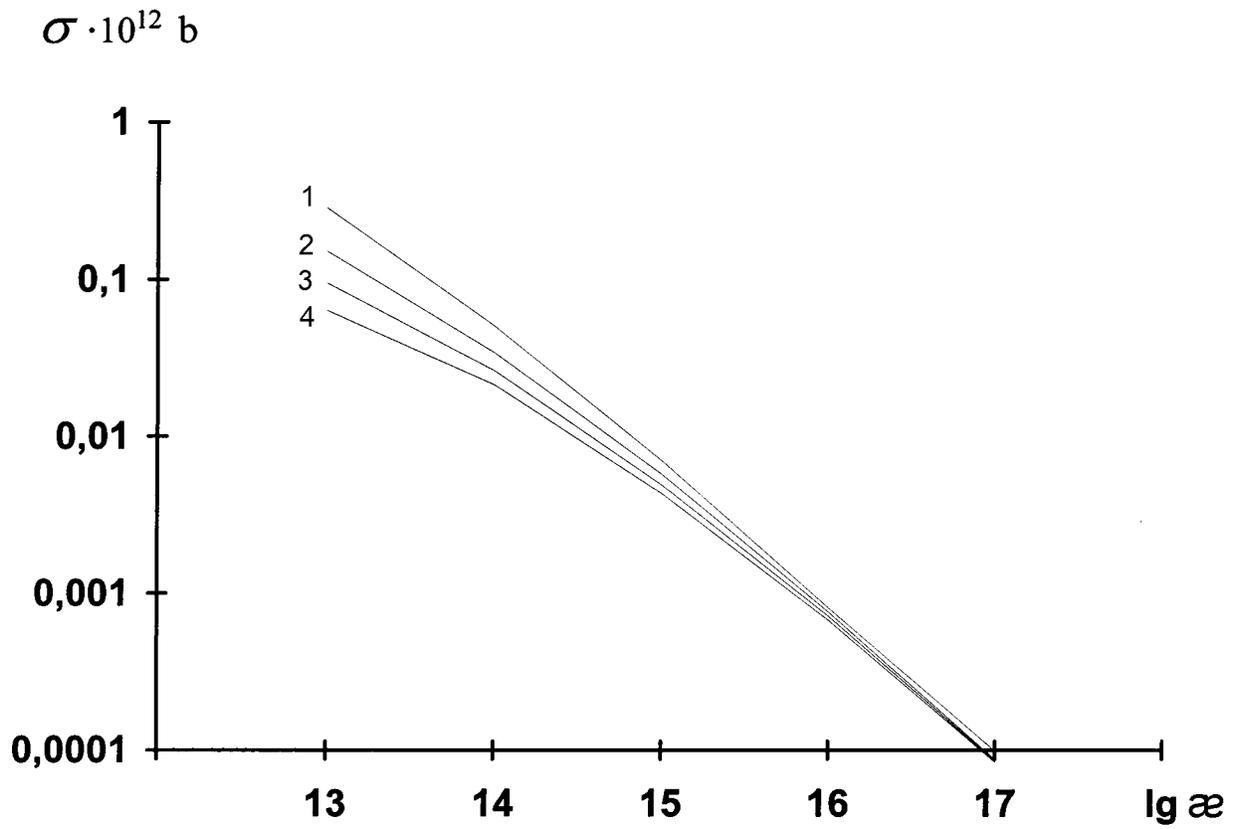}}
\caption{The   cross section of the process $e+\gamma \to e+Z+H$
as a function of $\ae$ for various values of the Higgs mass:
$M_H=100$ (1), 200 (2), 300 (3),\ \ \  400 (4) GeV}
\end{figure}
\bel{sigin}
\sigma(e^++e^-\to Z+H)=\frac{G^2_f M^4_Z}{48\pi s}
(1-4\sin^2\Theta_W+8\sin^4 \Theta_W),
\ee
whereas with consideration for  (\ref{sig})  the rate of the
Higgs boson production channel $e+\gamma \to e+Z+H$  in the logarithmic approximation
($\ln\frn{\ae m^2}{M^2}\gg 1$) is described as
$$
\sigma(e^-+\gamma \to e^-+Z+H)=\left\{\sigma_1,\; k \gg1, \atop 
\sigma_2,\; k \ll 1, \right.
$$
$$
k=\frac1{\ae}\left(\frac{ M_Z}{m} \right)^4 \ln\frac{\ae m^2}{M^2},
$$
\bel{log}
\sigma_1=\frac13(g_V-g_A)^2\left(\frac{eG_f m}{\pi}\right)^2
\frac1{\ae}\left(\frac{M_Z}{m}\right)^4 \ln\left(\frac{\ae m}{M}\right)
\ln\left(\frac{\ae m^2}{M^2}\right),
\ee
$$
\sigma_2=\frac23g^2_A \left(\frac{eG_f m}{\pi}\right)^2 \ln\frac{\ae m^2}{M^2}.
$$
 
	The study  of (\ref{sigin}), (\ref{log}) for the head-on electron--photon collisions when 
the energies of the particles are equal,
leads  to the following ratio of  the cross sections: 
$$
\frac{\sigma(e+\gamma \to e+Z+H)}{\sigma(e^++e^-\to Z+H)}\cong
\left\{C_1,\; k \gg1,\, 
 \, \ae\ll 10^{22}, \atop
C_2,\; k \ll 1,\, 
\, \ae\gg 10^{23},\right.
$$
\bel{phe}
C_1=5\alpha \ln\left(\frac{2E}{m}\right)\ln\left(\frac{4E^2}{M m}\right),\;
C_2=\alpha \ln\left(\frac{2E}{M}\right)\ae\left(\frac{m}{M_Z}\right)^4,
\ee
where $\alpha$ is the fine structure constant. 

The results obtained (\ref{log}), (\ref{phe}) are valid in the wide  range of  external field
intensities, and energy values. For example, if
$E>1000$ GeV, the ratio of the cross sections of the
processes  compared is equal to $C_1>10.$ 

	Thus, as it follows from (\ref{phe}), at least at high energies 
the cross section of the 
reaction 
examined can substantially exceed that of the process $e^++e^-\to Z+H,$ which at present is 
considered to be the most probable channel for the Higgs boson production.

\end{document}